\documentclass[useAMS,usenatbib,usegraphicx]{mn2e}
\usepackage{epsfig}
\usepackage{amsmath} 
\usepackage{rotating}           
\usepackage{color}     
\usepackage{graphicx}
\usepackage{times}
\usepackage{upgreek} 
\usepackage{multicol}

\def \HI {H{\sc \,i}}

\def\lapp{\ifmmode\stackrel{<}{_{\sim}}\else$\stackrel{<}{_{\sim}}$\fi}
\def\gapp{\ifmmode\stackrel{>}{_{\sim}}\else$\stackrel{>}{_{\sim}}$\fi}

\title[QSO photometric redshifts using machine learning]{QSO photometric redshifts using machine learning and neural networks}

\author[S. J. Curran, J. P. Moss \& Y. C. Perrott]{S. J. Curran\thanks{Stephen.Curran@vuw.ac.nz}, J. P. Moss and Y. C. Perrott\\
School of Chemical and Physical Sciences, Victoria University of Wellington, PO Box 600, Wellington 6140, New Zealand}

\begin{document}

 \date{Accepted ---. Received ---; in original form ---}

\pagerange{\pageref{firstpage}--\pageref{lastpage}} \pubyear{2021}

\maketitle
\label{firstpage}
\begin{abstract}
  The scientific value of the next generation of large continuum surveys would be greatly increased if the redshifts of
  the newly detected sources could be rapidly and reliably estimated.  Given the observational expense of obtaining
  spectroscopic redshifts for the large number of new detections expected, there has been substantial recent work on
  using machine learning techniques to obtain photometric redshifts. Here we compare the accuracy of the predicted
  photometric redshifts obtained from {\em Deep Learning} (DL) with the {\em $k$-Nearest Neighbour} (kNN) and the {\em
    Decision Tree Regression} (DTR) algorithms.  We find using a combination of near-infrared, visible and
  ultraviolet magnitudes, trained upon a sample of SDSS QSOs, that the kNN and DL algorithms produce the best
  self-validation result with a standard deviation of $\sigma_{\Delta z} =0.24$ ($\sigma_{\Delta z (\text{norm})}
  =0.11$). Testing on various sub-samples, we find that the DL algorithm generally has lower values of $\sigma_{\Delta
    z}$, in addition to exhibiting a better performance in other measures. Our DL method, which uses an easy to
  implement off-the-shelf algorithm with no filtering nor removal of outliers, performs similarly to other, more
  complex, algorithms, resulting in an accuracy of $\Delta z < 0.1$ up to $z\sim2.5$.  Applying the DL algorithm trained
  on our 70\,000 strong sample to other independent (radio-selected) datasets, we find $\sigma_{\Delta z} \leq0.36$
  ($\sigma_{\Delta z (\text{norm})} \leq0.17$) over a wide range of radio flux densities. This indicates much potential
  in using this method to determine photometric redshifts of quasars detected with the Square Kilometre Array.
\end{abstract}  

\section{Introduction} 
\label{intro}

Continuum surveys on the next generation of telescopes, e.g.  the {\em Square Kilometre Array}
(SKA), are expected to yield a large number of sources for which the redshifts are unknown.  Even
the {\em Evolutionary Map of the Universe} (EMU, \citealt{nha+11}) on the {\em Australian Square
  Kilometre Array Pathfinder}, an SKA precursor, is expected to yield 70 million distant radio
sources.  Given the observational expense of high quality spectroscopic data, there is currently
much activity in developing reliable photometry-based redshifts for distant sources (\citealt{lnp18}
and references therein).

The need for optical spectroscopy also causes a bias towards the most luminous sources, leaving the
more obscured, gas-rich objects undetected \citep{cwm+06,cwc+11}. For these, the redshifts would
ideally be obtained from the radio photometric properties, although this is proving difficult
\citep{maj15,nsl+19}, due to their relatively featureless spectral energy distributions (SEDs) and
the limited number of radio sources for which redshifts are available \citep{msc+15}.

The optical band photometry methods generally use machine learning models trained on the $u- g$, $g
- r$, $r - i$ \& $ i - z$ colours of the {\em Sloan Digital Sky Survey} (SDSS), upon which they are
self-validated (e.g. \citealt{rws+01,bbm+08}). Due to the large redshift ranges being explored,
\citet{cm19} noted the importance of the bands beyond the visible and, including the
WISE\footnote{{\em Wide-Field Infrared Survey Explorer} (\citealt{wem+10}).}  and
GALEX\footnote{{\em Galaxy Evolution Explorer} data release GR6/7 \citep{bst17}.}  colours as
features in the {\em $k$-Nearest Neighbour} algorithm, \citet{cur20} found a significant
increase in the accuracy over using the SDSS colours alone, bringing the standard deviation in 
$\Delta z \equiv z_{\rm spec} - z_{\rm phot}$ down from $\sigma_{\Delta z}[\text{data}] = 0.525$ to 0.314.

The $FUV - NUV$, $NUV-u$, $u- g$, $g - r$, $r - i$, $i - z$, $z-W1$ \& $W1-W2$ colours of the SDSS sample also provided
a suitable training set for another independent dataset of radio-loud sources (i.e. quasars, as well as the optically
selected QSOs).  In this paper, we use the full set of colours to compare other machine learning methods to the kNN,
specifically {\em Decision Tree Regression} and {\em Deep Learning}, and again explore their transferability in training
data for other, radio-selected, surveys, thus testing their potential in yielding reliable photometric redshifts for
sources detected in forthcoming radio continuum surveys.

\section{Methods and Results}

\subsection{The training data}
\label{sec:data}

We extracted the first 100\,000 QSOs with accurate spectroscopic redshifts ($\delta z/z<0.01$) from the SDSS Data
Release 12 (DR12, \citealt{aaa+15}). These were then matched to sources in the {\em NASA/IPAC Extragalactic Database}
(NED), with the NED names being used to scrape the {\em Wide-Field Infrared Survey Explorer} (WISE), the {\em Two Micron
  All Sky Survey} (2MASS, \citealt{scs+06}) and GALEX databases.  
As per \citet{cur20}, in order to 
ensure a uniform magnitude measure between the SDSS and other samples (Sect.~\ref{atod}), 
for each QSO the PSF flux densities associated with the AB magnitudes which fell within $\Delta\log_{10}\nu = \pm0.05$ of the 
central frequency of the band were added. 
Within each band range the fluxes were then averaged
before being converted to a magnitude.\footnote{For GALEX this was via 
$M = -2.5(\log_{10}S_{\nu}-3.56)$, where $S_{\nu}$ is the specific flux density in Jy (http://galex.stsci.edu/gr6/).}
This binning has the advantage of being applicable to
other samples for which the SDSS photometry may not be directly available, but where there is other nearby
photometry in other databases. 
That is, when using the SDSS to train other independent data sets (see Sect.~\ref{atod}).\footnote{In addition to 
yielding  any radio photometry (see Sect.~\ref{rfd}).}
\begin{figure}
\centering \includegraphics[angle=-90,scale=0.48]{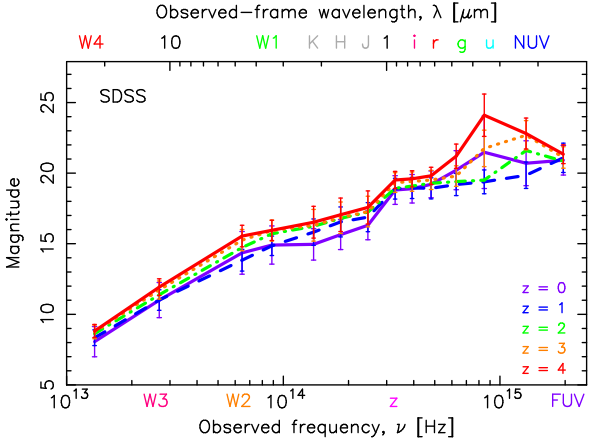}
\caption{The mean magnitudes of the SDSS QSOs at redshifts of $z< 0.2$, $0.95 < z< 1.05$, $1.95
  < z< 2.05$, $2.8 < z< 3.2$ and $3.5 < z< 4.5$. 
The positions of the mid-infrared to ultra-violet  photometric bands are indicated along the top and bottom axes.
The error bars show  $\pm1\sigma$ from the mean.
The inversion of the trend  at $\nu\gapp10^{15}$~Hz, where the ultra-violet fluxes are expected to be generally undetectable,
 confirms our suspicion that there is confusion arising at the magnitude extremes.}
\label{SED}
\end{figure}
The mean magnitudes close to redshifts of $z=0,1,2,3$ and 4 of the sample are shown in 
Fig.~\ref{SED}. Retaining the sources detected  in all nine bands
($FUV,NUV,u,g,r,i,z,W1,W2$) left a sample size  71\,267 QSOs, which is 71\% of the original data.

\subsection{Algorithms}

As previously \citep{cur20}, we use the $FUV-NUV$, $NUV-u$, $u- g$, $g - r$, $r - i$, $ i- z$,
$z-W1$ \& $W1-W2$ colours, in addition to the $r$ magnitude, as features. As per typical practice,
we train on 80\% of the data and validate on the remaining 20\%, quantifying the result via the
standard deviation of the photometric redshifts from the spectroscopic. That is,
\[
\sigma_{\Delta z} = \sqrt{\frac{1}{N}\sum_{i=1}^N \Delta z^2},
\]
which we give for both the data and the Gaussian fit, since the latter is quoted in some of the literature (generally, $\sigma_{\Delta z}[\text{fit}]<\sigma_{\Delta z}[\text{data}]$).
Also quoted is the {\em normalised standard deviation}  (e.g. \citealt{dp18,lnp18}), which 
is obtained from 
\[
\Delta z (\text{norm}) \equiv \frac{z_{\rm spec} - z_{\rm phot}}{z_{\rm spec} +1},
\]
giving $\sigma_{\Delta z (\text{norm})} < \sigma_{\Delta z}$ for the same data. Finally, we also give the {\em median absolute deviation}
(MAD),
\[
\sigma_{\text{MAD}} \equiv 1.48 \times \text{median}\left|z_{\rm spec} - z_{\rm phot}\right|,
\]
and the 
{\em normalised median absolute deviation} (NMAD),
\[
\sigma_{\text{NMAD}} \equiv 1.48 \times \text{median}\left|\frac{z_{\rm spec} - z_{\rm phot}}{z_{\rm spec} +1}\right|,
\]
e.g. \citet{bcd+13} and \citet{asu+17}, respectively. 
Note that, due to
the randomisation of the training and validation data, the values of $\sigma$ exhibit
some slight variation about the values quoted.

\subsubsection{$k$-Nearest Neighbour}
\label{kNN}

The kNN algorithm, which compares the Euclidean distance between a datum and its $k$ nearest neighbours in a feature
space, has had some success in predicting photometric redshifts. However, the standard practice of using the $u- g$, $g
- r$, $r - i$ and $ i - z$ colours alone results in poor predictive power at $z\gapp2$ and non-Gaussian distributions of
$\Delta z$ \citep{rws+01,wrs+04,mhp+12,hdzz16}, with \citet{cur20} obtaining $\sigma_{\Delta z}[\text{data}] = 0.525$.
By including the GALEX colours as features in the {\tt KNeighborsRegressor} function of {\sf
  sklearn}\footnote{https://scikit-learn.org/stable/}, this was reduced to $\sigma_{\Delta z}[\text{data}] = 0.352$,
with the addition of near-infrared (NIR) $W1$ \& $W2$ bands bringing this down to $\sigma_{\Delta z}[\text{data}] =
0.314$ \citep{cur20}. Note that the addition of the mid-infrared (MIR) $W3$ \& $W4$ bands did not have any noticeable
benefit, while reducing the fraction of sources with detections in all of the bands from 79\% to 57\% of the SDSS
sample.

The addition of the NIR bands tightened the $z_{\rm phot} - z_{\rm spec}$ correlation over all redshifts, probably due
to these spanning $\lambda\sim1~\mu$m inflection in the SEDs (see Sect.~\ref{physint}), and inclusion of the UV bands
had the most profound effect at low redshifts, probably due to sampling of the Lyman-break (see Sect.~\ref{exbu}).
Inclusion of other bands had been previously explored, giving similar results
(e.g. \citealt{bbm+08,bmh+12,bcd+13,ywf+17,dbw+18,sih18}).  Re-testing the kNN algorithm on our larger dataset, 71\,267
cf. 26\,301, we find significant improvement with $\sigma_{\Delta z}[\text{data}] = 0.236$ (Fig.~\ref{f1}, left),
although, in addition to the larger sample, some of this will be due to the 80:20 training to validation ratio,
cf. 50:50 \citep{cur20}.

\subsubsection{Decision Tree Regression}
\label{DTR}

Another common method used to determine photometric redshifts is the DTR algorithm, which builds a regression model in
the form of a tree structure, branching a large dataset into smaller subsets starting from the entire dataset (the top
node).  This branches into two child nodes, based on a predefined decision boundary, with one node containing data above
the decision boundary and one below \citep{icvg14}. This decision-based bifurcation continues recursively until a
predefined stopping criterion is reached. In this case, a certain value of $\Delta z$.  Using the full set of colours as
features for the {\tt DecisionTreeRegressor} function of {\sf sklearn}, we find a maximum tree depth of $\approx10$ to
be optimal. Values greater than this result in over-fitting, making the $\Delta z$ distribution less Gaussian with a
wider spread in $\sigma_{\Delta z}[\text{data}]$ (although $\sigma_{\Delta z}[\text{fit}]$ narrows considerably). With
$\sigma_{\Delta z}[\text{data}]=0.333$ (Fig.~\ref{f1}, middle), the DTR algorithm does not appear to perform as well as the kNN.

\subsubsection{Deep Learning} 

The concept of deep learning is to configure a computer architecture based upon the natural neurons found in biological
brains, thus being very flexible, non-linear and sensitive to patterns in multi-dimensional data.  The artificial neuron
(``perceptron'') is the functional unit of the DL algorithm and consists of several layers: an input $x_i$; weights and
biases, $w_i$, defining the importance of the inputs; an activation function which transforms the weighted input -- this
is usually a non-linear function, such as the {\em hyperbolic tangent} (tanh), a {\em sigmoid function}, or the more
step-wise {\em Rectified Linear Unit} (ReLu) function.  Each perceptron of the input layer may be connected to one, more
or all of the other perceptrons in the adjacent layers, and the inputs are multiplied by the weights. The output then
becomes
\[
y_{k}=g\left(\sum_{j=0}^{M} w_{k, j}^{(N-1)} z_{i}^{(N-2)}\right),
\]
where $z_i$ is the output of the \textit{i}th perceptron in the $N-1$ level \citep{tlm+03}. 

We use {\sf TensorFlow}\footnote{https://www.tensorflow.org}, an off-the-shelf deep learning library. Since our output
layer is an estimate of the redshift, the learning is supervised and, after testing various combinations of
hyperparameters, we found a simple model with two {\em ReLu} layers and one {\em tanh} layer comprising 200 neurons each
(giving 82\,601 trainable parameters) to be the most effective, with larger and additional layers being of little
benefit while considerably increasing the computation time. Since performing a regression, we employed the {\sf keras}
{\tt RMSprop} optimizer, using early stopping with a patience of 10 to ensure against over-fitting.

\begin{figure*}
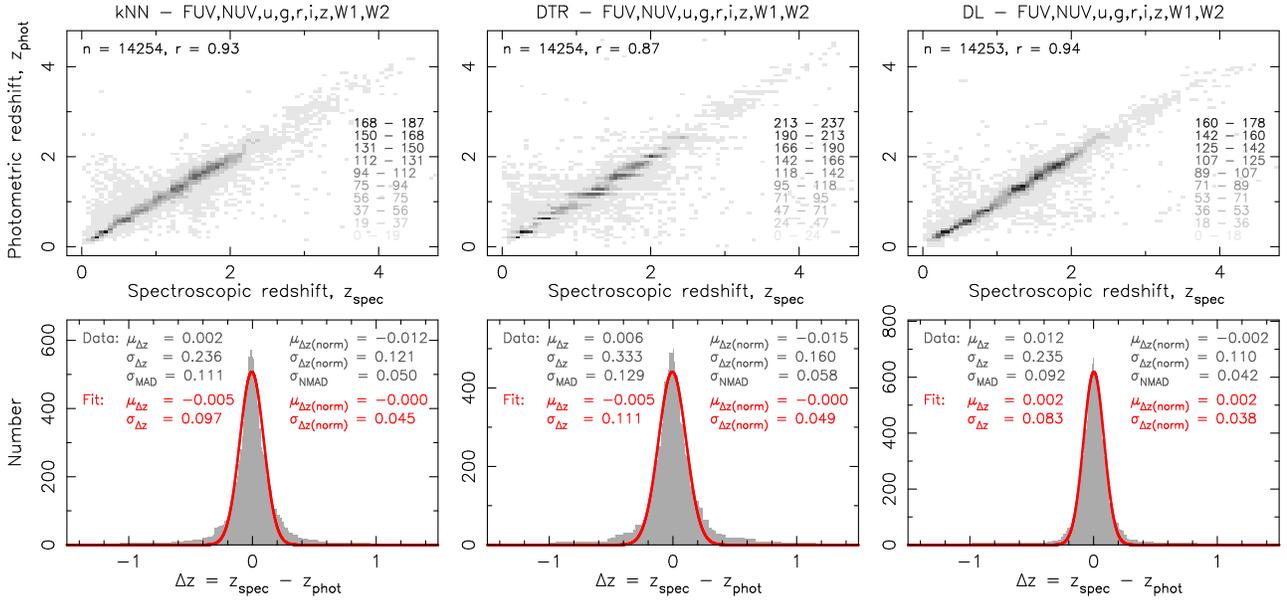

\setlength{\lineskip}{10pt}
\centering \includegraphics[angle=-90,scale=0.37]{kNN-train_0.8.eps}
\centering \includegraphics[angle=-90,scale=0.37]{DTR-5-max_depth=10.eps}
\centering \includegraphics[angle=-90,scale=0.37]{TF-colours_200-train_0.8.eps}
\caption{The predictions based on the colours and $r$ magnitude from the different methods (kNN -- left, DTR -- middle,
  DL -- right) using the SDSS sample, from a 80:20 training-validation split. In the scatter plot $r$ shows the
  regression coefficient and the numbers to the right show the number of sources within each grey-scale pixel. The
  histogram shows the $\Delta z$ distribution, where the mean and standard deviation are given for both the distribution
  and the Gaussian fit.}
\label{f1}
\end{figure*}

As with the kNN and DTR methods, we use the standard  80:20 training--validation split, shuffling the data.  
Using the same colour
combinations as the other algorithms, we find that the DL algorithm provides 
predictions of similar accuracy to the kNN method with $\sigma_{\Delta
  z}[\text{data}]=0.236$ (Fig.~\ref{f1}, right).
We note that the
using the magnitudes as features in the DL algorithm, rather than the colours, gives a very similar result.
\citet{bcd+13,pp18} and \citet{bsf+20} also use magnitudes as features for their algorithms and we interpret this as the
neutral net being free to decide how best to use this more elemental data.

\subsection{Comparison of the algorithms}
\label{coa}

\begin{figure*}
\setlength{\lineskip}{10pt}
\centering \includegraphics[angle=-90,scale=0.55]{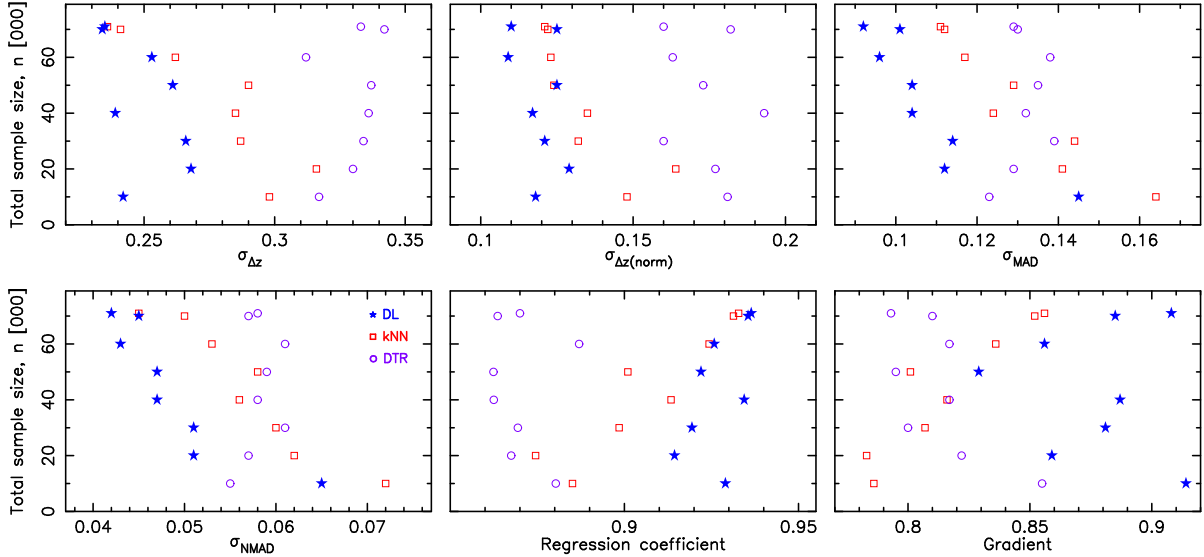}
\caption{The total sample size versus the performance of each algorithm, as measured by the standard deviation,
  $\sigma_{\Delta z}$, the normalised standard deviation, $\sigma_{\Delta z (\text{norm})}$, the median absolute
  deviation, $\sigma_{\text{MAD}}$, the normalised median absolute deviation, $\sigma_{\text{NMAD}}$, the regression
  coefficient and the gradient of the linear regression between $z_{\rm phot}$ and $z_{\rm spec}$. The kNN algorithm is
  represented by unfilled squares, the DTR by unfilled circles and the DL by filled stars. The DL algorithm has the best performance as judged by all indicators.}
\label{fcoa}
\end{figure*}

In Fig.~\ref{fcoa}, we show the performance of each of the algorithms for various sample sizes, using the same 80:20
training-validation split.
From the standard deviation, we see that the DL algorithm generally performs best and, like 
the kNN, exhibits the expected anti-correlation between $\sigma_{\Delta z}$ and sample size.  There is some
scatter, most notably the DL point at $\sigma_{\Delta z}\approx0.24$ \& $n=10\,000$, although this is to be 
expected due to the shuffling of the training and validation data
in these single experiments. For the DTR algorithm, there is no apparent
relationship  between $\sigma_{\Delta z}$ and $n$, which also appears to be case for all other measures.
The superior performance of the DL method is confirmed by the other measures, where this generally has 
lower values of $\sigma_{\Delta z (\text{norm})}$, $\sigma_{\text{MAD}}$ and  $\sigma_{\text{NMAD}}$, while 
having higher values of the regression coefficient and a gradient which is generally closer to unity than
the kNN and DTR algorithms.\footnote{These tend to be ``dragged''  down from unity by the excess of 
$z_{\rm spec}\gg z_{\rm phot}$ points at $z_{\rm spec}\gapp3$ (see Sect.~\ref{cur_lim}).}

\section{Discussion}
\label{disc}

\subsection{Comparison with previous deep learning results}
\label{cwpm}

Testing the three popular methods, we find deep learning to be the best in terms of self-validation.  Comparing with
other DL photometric redshift estimates of QSOs, \citet{ldlr11} used the {\em Weak Gated
  Experts} method, which utilises a feature space ($z_{\rm phot}$) which is partitioned by $FUV-NUV$, $NUV-u$, $u- g$,
$g - r$, $r - i$, $ i- z$ colour, and an {\em expert} maps each pattern of feature space to a target ($z_{\rm spec}$),
with the output of the experts defining a new feature space.  This resulted in a sample of $\approx27\,000$ from the
original 105\,783 QSOs of the SDSS DR7 \citep{srh+10}.  Using a 60:40 training:validation split, the gate network uses a
\textit{softmax} activation function to map extracted patterns from the new feature space to the target space, which is
an extension of the original feature space, with the added expert predictions. With this method they obtain
$\sigma_{\Delta z} [\text{fit}] =0.198$ and $\sigma_{\text{NMAD}} = 0.029$ compared to our 0.083 and 0.042,
respectively.

\citet{bcd+13} used the $u,g,r,i,z,NUV, FUV, W1,W2,W3,W4$, in addition to the UKIDSS DR9 $Y,J,H,K$ near-infrared bands
in a {\em Multilayer Perceptron Quasi-Newton Algorithm}, in which the Hessian of the error function finds the local
minima and maxima of the activation function's scalar error field.  They use a 60:40 training:testing split, where
10-fold cross validation is performed on the training set.  Using all of the SDSS, WISE, GALEX \& UKIDSS bands, they
obtain $\sigma_{\Delta z}=0.15$ and $\sigma_{\text{MAD}} = 0.040$, outperforming our model (Fig.~\ref{f1},
right). However, this does require six additional bands, giving just 13\% of the full sample, cf. 71\% for ours.
This would also lead to similar reductions in sample size when applied to other sources (see Sect.~\ref{atod}).

\citet{dp18} combined a {\em Mixture Density Network} and a {\em Deep Convolution Network} into a hybrid architecture,
which they term a {\em Deep Convolutional Mixture Density Network}. Like us, they used a rectified linear unit applied
to multiple layers, although they extracted the features from the SDSS DR7 \& DR9 images. These comprise all five SDSS
magnitudes and their colour combinations, although we find no extra benefit by combining both magnitudes and colours for
the SDSS-W1-W2-GALEX data.  The image-based algorithm automatically selects the best features, which required cluster
computing to run.  Using 185\,000 sources from the SDSS DR7 and DR9 \citep{srh+10,ppa+12}, from their 54:46
training:validation ratio, they obtained $\sigma_{\Delta z} [\text{fit}] = 0.217$, $\sigma_{\Delta z
  (\text{norm})}[\text{fit}]=0.095$.  and $\sigma_{\text{NMAD}}= 0.026$.

\citet{pp18} used a {\em Convolutional Neural Network} (CNN) to detect and classify QSOs  from the light curves 
in SDSS data (Stripe 82).  Their architecture was based upon a three-layer CNN, which convolved the raw input with
either a temporal convolution (time dependent magnitudes) or a filter convolution (integrated magnitudes).  They used a
hyperbolic tangent activation function applied to three layers and  classified with an 80:20 training:validation ratio.
Their architecture gives the probability of a signal being in a particular redshift range and, using
the normalised root mean square,
they found 80.4\% of the photometric redshifts within $\Delta z (\text{norm}) < 0.1$, 87.1\% within $ \Delta z
(\text{norm}) < 0.2$ and 91.8\% within $\Delta z (\text{norm}) < 0.3$, compared to our 91.0\%, 96.6\% and 98.0\%,
respectively.  Comparing their results with the kNN, support vector machine, a random forest classifier and a Gaussian
process classifier, they found that their method compared most favourably to the kNN, especially for $z < 2.5$. The low
number of quasars at higher redshifts means that their method was not trained well at $z\gapp2.5$.

Similar to us, \citet{bsf+20} use {\sf TensorFlow} with ReLu activation functions, but with all three-layers using 512
neurons and the \textit{Adam} optimizer, upon the Pan-STARRS1 $g,r,i,z,y$ magnitudes. Using an 80:20 
training:validation split, they achieve a simulated $\Delta z (\text{norm}) [\text{fit}]= 0.032$
and  $\sigma_{\text{NMAD}} = 0.016$ over a 
limited $z_{\rm spec}\approx0-1$,
acknowledging that this does not fully capture the variance in the data and includes galaxies. This redshift limitation
permits reasonable predictions over such a small number of bands, due to minimisation of the shift in the 
 rest-frame bands (Sect.~\ref{physint}), and is likely the reason for the more accurate predictions of
galaxy, cf. QSO, redshifts (e.g. \citealt{bbm+08,ldlr11,bcld14,dp18,ach+19}). 

We summarise these in Table~\ref{C1}, where we see that our DL method produces similar results to the literature.
\begin{table*}
\centering
  \caption{Comparison with other deep learning methods applied to QSOs. $n$ gives the total number of sources (test + validation). 
    Note that $\sigma_{\Delta z (\text{norm})} [\text{data}]$ is only quoted by \citet{bcd+13} ($\sigma_{\Delta z (\text{norm})} [\text{data}]=0.15$).
    The radio samples for this paper utilise the SDSS DR12 DL model (Sect.~\ref{atod}).}
\begin{tabular}{@{}l  l   c l c r  c  c r @{}} 
\hline
\smallskip
Reference& Sample &Photometry & $z$-range & $n$ &\multicolumn{2}{c}{Data} & \multicolumn{2}{c}{Fit}\\
                   &               &     & &  &  $\sigma_{\text{NMAD}}$& $\sigma_{\Delta z (\text{norm})}$  & $\sigma_{\Delta z}$  & $\sigma_{\Delta z (\text{norm})}$ \\
\hline
\citet{ldlr11} &   SDSS DR7 &  $FUV, NUV,u,g,r,i,z$&   $\lapp4$    &    27\,000   &   0.029&   --- &0.198  &   --- \\
\citet{bcd+13} &   SDSS DR7     &  15 bands  &      $\leq3.6$             &   14\,000    & --- & 0.069 &  --- & ---\\
\citet{dp18} & SDSS DR9 &  $u,g,r,i,z$  &  $\lapp5$  & 185\,000&  0.026 & --- & 0.217 & 0.095\\
\citet{pp18} &  SDSS DR7 (Stripe 82)  & $u,g,r,i,z$  & $\lapp2.6$& 9000    &   --- & 0.349 & --- & ---\\
\citet{bsf+20} & Pan-STARRS1 & $g,r,i,z,y$ & $\lapp1$  & 540\,244&  $>0.016$ &  --- &  --- & $>0.032$\\
This paper  & SDSS DR12 &  9 bands&  $\leq4.9$   & 71\,267  & 0.042& 0.110  & 0.083 &  0.038\\
                & OCARS &  ... &  $\leq4.8$   &  649 & 0.063 & 0.170 & 0.107 & 0.050 \\
                & LARGESS & ... &  $\leq4.7$   &  1046 & 0.057 & 0.123 & 0.102 & 0.049 \\
                & FIRST & ... &  $\leq4.4$   & 6129 & 0.048 & 0.110 & 0.089 & 0.042 \\
                & 21-cm & ... & $\leq3.6$   &  28 & 0.124 & 0.140 & 0.165 & 0.093 \\
\hline   
\end{tabular}
\label{C1}  
\end{table*} 
However, unlike these other methods, which use magnitudes directly and validate on the same database,
our scraping of the photometry allows us to apply the SDSS training to other (radio selected) samples, giving comparable results
(Sect.~\ref{atod}).

\subsection{Transferability to other surveys}
\label{atod}

Our goal is to obtain statistical redshift estimates for the radio continuum source surveys to be undertaken with the
SKA and its pathfinders.  Given this, we are interested in whether our methods are transferable to other
(radio-selected) data, although, as outlined in \citet{cm19}, finding such datasets of sufficient size for which
redshifts are available is a challenge. One dataset is the {\em Optical Characteristics of Astrometric Radio Sources}
(OCARS) catalogue of {\em Very Long Baseline Interferometry} astrometry sources \citep{mal18}, a sample of flat spectrum
radio sources with $S$-band (2--4~GHz) flux densities ranging from 15~mJy to 4.0~Jy \citep{mab+09}.

Applying the kNN training of the full set of colours to
the OCARS sources gave $\sigma_{\Delta z}[\text{data}] =0.356$ for the 739 of the 3663 sources which had all of the
required photometry \citep{cur20}.  Here, we perform the same test, but using the apparently superior DL model, in
addition to testing this on the {\em Large Area Radio Galaxy Evolution Spectroscopic Survey} (LARGESS).
\begin{figure}
\centering \includegraphics[angle=-90,scale=0.52]{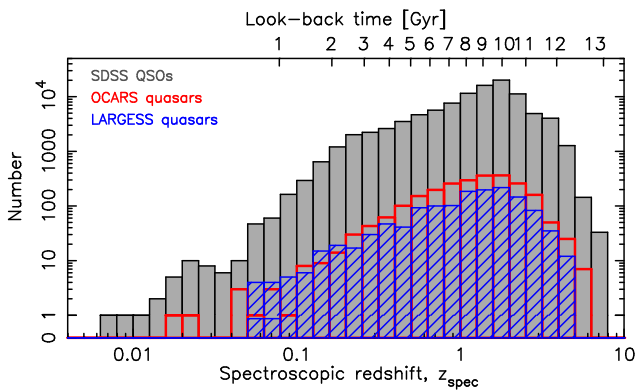}
\caption{The redshift distribution for the SDSS (filled) and OCARS (unfilled) and LARGESS (hatched histogram) QSOs/quasars.}
\label{z_QSO}
\end{figure}
Although both OCARS and LARGESS are independent, we see that the redshift distributions are well matched (Fig.~\ref{z_QSO}).

We also test a sample from the {\em Faint Images of the Radio Sky at
  Twenty-Centimeters} (FIRST, \citealt{bwh95,wbhg97}) survey, from
which the redshifts have been obtained by matching with SDSS DR14 QSOs \citep{ppa+18}. Although, arguably
not as independent a sample as OCARS or LARGESS, this provides 18\,273 radio sources with redshifts with 
which to test the model.

\subsubsection{OCARS}

\begin{figure*}
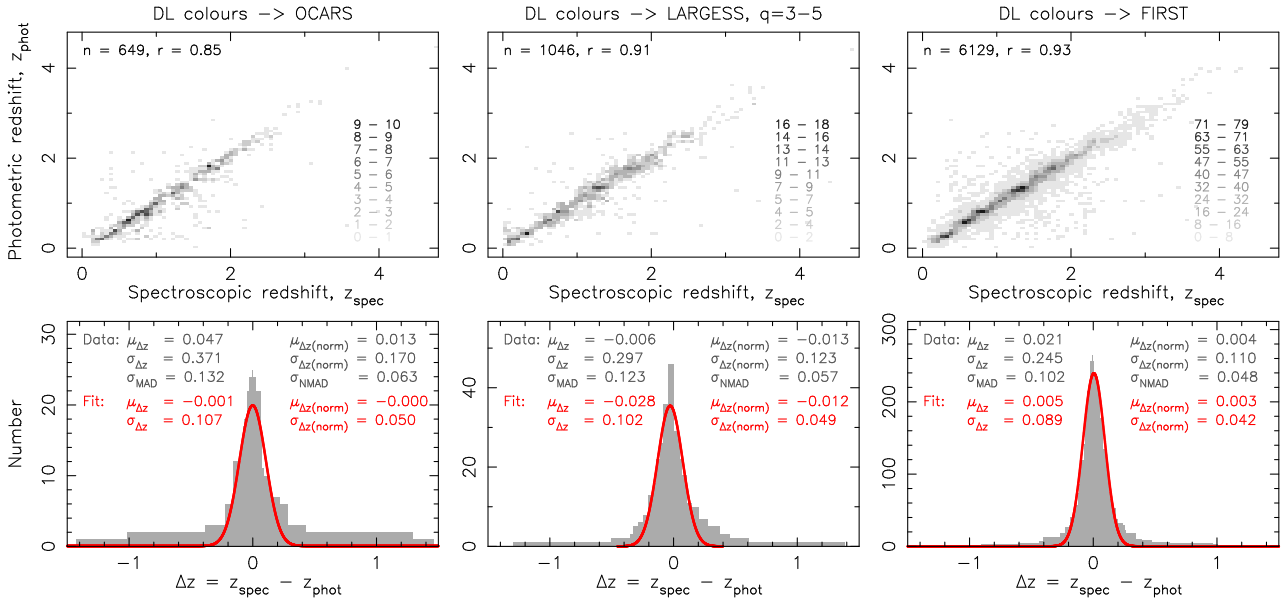

\setlength{\lineskip}{10pt}
\centering \includegraphics[angle=-90,scale=0.37]{X-OCARS_colours.eps}
\centering \includegraphics[angle=-90,scale=0.37]{X-LARGESS_q=3-5_colours.eps}
\centering \includegraphics[angle=-90,scale=0.37]{X-FIRST_colours_1.eps}
\caption{The predictions for the other samples trained using the colours from the DL training on 
SDSS data and validation on the OCARS data (left), the  LARGESS $q=3-5$ data (middle) and the 
FIRST data (right). Using the $FUV,NUV,u,g,r,i,z,W1,W2$ magnitudes directly gives similar results.
}
\label{X2}
\end{figure*}

\begin{figure}
\centering \includegraphics[angle=-90,scale=0.48]{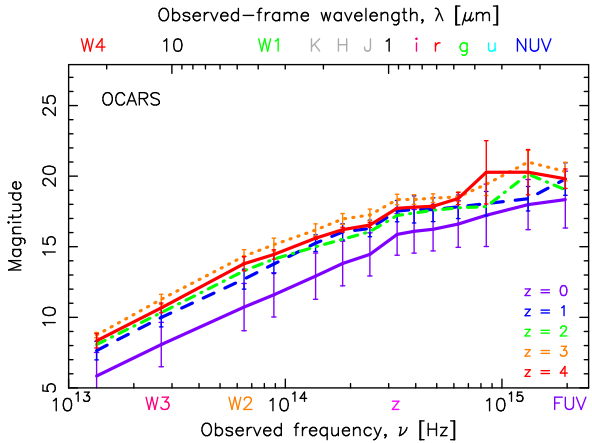}
\caption{The mean magnitudes of the OCARS quasars at redshifts of $z< 0.2$, $0.95 < z< 1.05$, $1.95
  < z< 2.05$, $2.8 < z< 3.2$ and $3.5 < z< 4.5$. The error bars show  $\pm1\sigma$ from the mean.
Again, we see the same inversion for the high redshift FUV magnitudes, indicating their unreliability (cf. Fig.~\ref{SED}).}
\label{OSED}
\end{figure}

Applying the DL training to the OCARS data, we find $\sigma_{\Delta z}[\text{data}] =0.362$
(Fig.~\ref{X2}, left), which is similar to the standard deviation obtained previously, using the kNN training.
From the SEDs (Fig.~\ref{OSED}), we see a similar pattern to the SDSS (Fig.~\ref{SED}), although
the OCARS sources are generally brighter, across all magnitudes, especially at $z\sim0$. We also note that the 
 mean NIR band for the OCARS sources is relatively featureless.

\subsubsection{LARGESS}

LARGESS
contains 10\,856 optical redshifts for 12\,329 radio sources \citep{csc+17} and comprises a mix of
galaxies and quasars. Of the latter,  
there are 1608 for which the redshift reliability flag is $q\geq3$ and 409 where $q=5$.\footnote{The redshift reliability flags range from $q=0-5$
by increasing quality, where, for example,  $q=0$ -- 
designates a ``poor-quality (or missing) spectrum'', $q=3$ -- ``a reasonably confident redshift'' and 
$q=5$ -- an ``extremely reliable redshift from a good-quality spectrum''.} Of these, there are 1046 and 292 QSOs
with all nine magnitudes, respectively. From Fig.~\ref{X2} (middle), we see that the
DL training on the SDSS sample provides an excellent prediction of the photometric redshifts,
even for the $q < 5$ sources with $\sigma_{\Delta z}[\text{data}] =0.297$. Limiting this to the $q=5$ quasars only
improves the prediction somewhat, giving  $\sigma_{\Delta z}[\text{data}] =0.217$, 
$\sigma_{\Delta z (\text{norm})}[\text{data}] =0.107$, $\sigma_{\text{MAD}}=0.102$ and $\sigma_{\text{NMAD}}=0.052.$
\begin{figure}
\centering \includegraphics[angle=-90,scale=0.48]{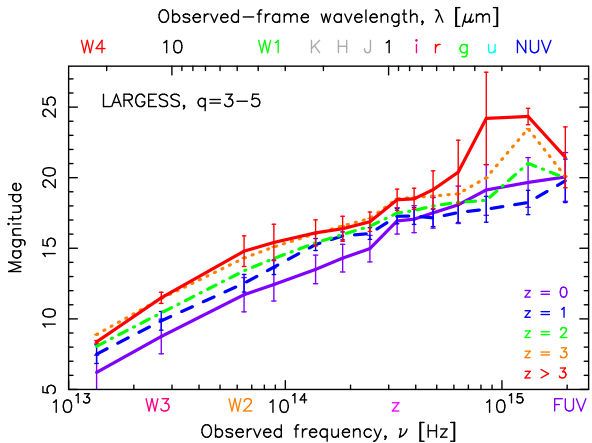}
\caption{The mean magnitudes of the LARGESS $q=3-5$ quasars  at redshifts of $z< 0.2$, $0.95 < z< 1.05$, $1.95
  < z< 2.05$, $2.8 < z< 3.2$ and $z> 3.2$ (due to a lower number of high redshift sources, Fig.~\ref{z_QSO}).
The error bars show  $\pm1\sigma$ from the mean. Again, we see the same inversion for the high redshift FUV magnitudes, indicating their unreliability (cf. Figs.~\ref{SED} \& \ref{OSED}).} 
\label{LSED}
\end{figure}
The SEDs bear more similarity to the SDSS QSOs than the OCARS quasars (Fig.~\ref{LSED}). This could be due to the LARGESS sources having been matched with SDSS counterparts \citep{csc+17}.

\subsubsection{FIRST}

The FIRST sample comprises 18\,273 quasars common to the SDSS DR14. Although the 
full photometry is available \citep{ppa+18}\footnote{https://www.sdss.org/dr14/algorithms/qso\_catalog/}, we obtain the data 
as per the other samples, in order to ensure its compilation in a consistent manner. 
In particular, the GALEX data compiled in \citeauthor{ppa+18} have been
force-photometered, recovering low signal-to-noise measurements not detected by GALEX. 
This, of course, leads to some matches not present in the GALEX GR6/7 catalogue and some extremely
high magnitudes, reaching $NUV = 35.2$ 
and $FUV = 37.6$.\footnote{Of the 219\,708 sources, 12\,487 have $NUV>25$ and $34\,872$ have  $FUV>25$, cf. Fig.~\ref{SED}.}
From our matching, 
6129 of the sample have all nine magnitudes, 
and, as may be expected, the quality of the  photometric redshifts are close to those of the 
SDSS validation data (Fig.~\ref{f1}, right).

\subsubsection{The 21-cm sample}

Upon considering the scarcity of a large catalogue of 
radio sources with spectroscopic redshifts (cf. \citealt{dwf+97,jws+02,bbp+08,ceg+17}),
\citet{cm19} tested their combination of magnitudes on the ``21-cm sample'', a compilation of radio sources
searched for associated \HI\ 21-cm absorption (\citealt{chj+19} and references therein). This comprises 819
radio-loud sources of known redshift, spanning  $0.002 \leq z \leq 5.19$
Of this, only 262
are quasars of which only 28 have the full magnitude compliment.
\begin{figure}
\centering \includegraphics[angle=-90,scale=0.42]{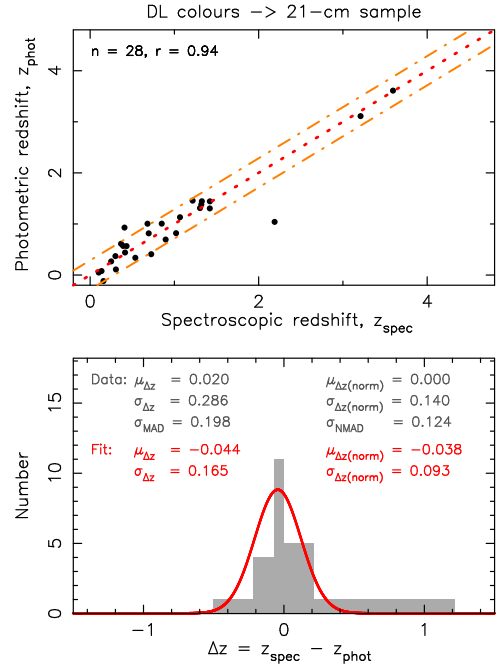}
\caption{The photometric redshift prediction for the 21-cm sample. The dotted line shows $z_{\rm phot} = z_{\rm spec}$ and
the broken lines $\pm\sigma_{\Delta z}[\text{data}]$.}
\label{f21}
\end{figure} 
Nevertheless, we apply the DL training on the SDSS sample to predict the redshifts (Fig.~\ref{f21})
and see that the photometric redshifts are reasonable for this small sample.

\subsubsection{Radio flux differences}
\label{rfd}

Given the improved sensitivities, 
many of the sources detected by the SKA and its pathfinders are expected to have flux densities many
times lower than the radio sources discussed here. 
For instance, EMU is expected to be sensitive to fluxes
of $S_{\rm radio} \lapp0.1$~mJy over large areas of the sky \citep{nha+11}, whereas, being VLBI calibration sources, 
OCARS sources have fluxes $S_{\rm radio} \gapp0.1$~Jy. 
As mentioned above, the requirement 
of a catalogue of radio sources with spectroscopic redshifts limits the number of independent 
test samples. However,  we should bear in mind that the training sample itself, although
optically selected, contains some radio sources. Of these, 3505 have at least one radio detection with the others either
being unsearched or with fluxes below the sensitivity of current surveys. 

Our mining of the source photometry
(Sect.~\ref{sec:data}) also includes radio data and, in order to quantify the radio fluxes, we extract the
$\nu<10$~GHz values. Where there is more than one flux for a given source we average these.
\begin{figure}
\centering \includegraphics[angle=-90,scale=0.48]{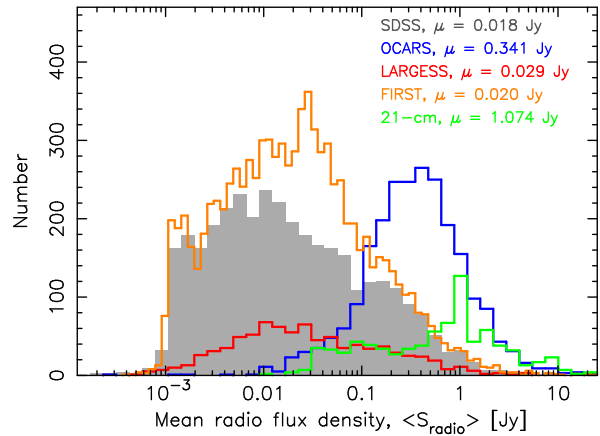}
\caption{The mean radio flux densities of the QSOs/quasars in the SDSS, OCARS, LARGESS ($q=3-5$),
FIRST and 21-cm  samples with measured $\leq10$~GHz flux densities. The legend shows the 
mean value of each distribution. For consistency, the 
fluxes for the FIRST sources are obtained in the same way as the other samples and the binning is
twice as fine as for the other samples in order to fit within a similar range. The 
1.4~GHz flux densities given in the FIRST catalogue range from 0.75~mJy to 14.8~Jy for this sample.}
\label{radio_hist}
\end{figure}
Showing the flux distributions in Fig.~\ref{radio_hist}, we do see that the OCARS sources are much brighter than the
SDSS, which in turn have a similar distribution to the LARGESS quasars, which are considered radio sources
\citep{csc+17}. We note that the  OCARS distribution appears more symmetrical than those of the SDSS and LARGESS, which
is suggestive of the weaker sources being truncated by the flux limited nature of the surveys. 
Due to small numbers, it is difficult to comment for the 21-cm quasars, although it is clear that the 
fluxes are about two orders of magnitude greater than the SDSS and LARGESS QSOs.
Given the results of testing on both OCARS and LARGESS (Fig.~\ref{X2}), as well as the 21-cm sample (Fig.~\ref{f21}),
we are confident that our algorithm can yield accurate photometric redshifts for  sources with 
a wide range of radio flux densities.

\subsubsection{Optical SEDs and radio loudness}

\begin{figure}
\centering \includegraphics[angle=-90,scale=0.51]{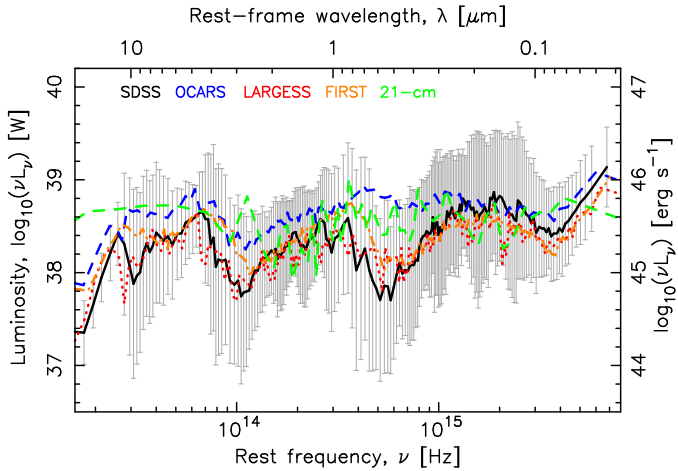}
\caption{The mean SEDs of the various sub-samples. The error bars show the $\pm1\sigma$
uncertainty in  the binned SDSS luminosities.}
\label{mean_SEDs}
\end{figure}

In addition to the differences in radio flux, it is known that the mean optical and IR SEDs
can differ between radio-loud and radio-quiet sources \citep{ewm+94}. In Fig.~\ref{mean_SEDs}, we show the
mean SEDs
from which  we see that the LARGESS and SDSS are very similar, which is not surprising given their similarity in mean
radio flux densities. As expected, the much more radio-loud OCARS and 21-cm quasars exhibit a  substantial difference
from the radio-quiet QSOs,
particularly in the optical band. The SEDs do, however, overlap with the SDSS $1\sigma$ uncertainties, which is in line
with the considerable variation between individual sources of the similar loudness noted by \citeauthor{ewm+94}.  Also,
the fact that the SDSS training applied to the other samples returns reasonable redshift predictions, suggests 
that the differences are small enough not to deviate significantly  from the trained model, although
they could contribute to the larger fraction of poor predictions for the OCARS data.

\subsubsection{Matching coincidence}

While the above does indicate that training the algorithm on the SDSS sources can provide reliable photometric redshifts
for the radio-selected samples, the requirement of the $u,g,r,i,z,W1,W2,NUV,FUV$ magnitudes does mean that for a
significant fraction of sources we cannot obtain a photometric redshift to the same reliability. Of the 1817 of the 3033
OCARS quasars which have at least one SDSS magnitude, 1136 have all five magnitudes, with 1034 of these also having 
$W1$ \& $W2$, and 649 (21\%)
having all of the required photometry.  For LARGESS, 1046 out of 1608 $q\geq3$ quasars have the full magnitude
complement, giving a matching coincidence of 65\%. 
\begin{figure}
\hspace*{-0.5cm}
\centering \includegraphics[angle=-90,scale=0.32]{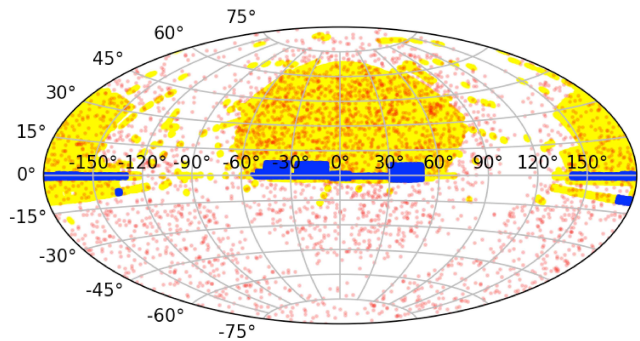} 
\vspace*{-1.0cm}
\caption{Right ascension and declination of the SDSS (northern sky -- yellow), OCARS (all sky -- red) and LARGESS ($-19^{\circ} < \delta < 8^{\circ} $ -- blue) sources.}
\label{sky}
\end{figure}
The limited OCARS--SDSS coordinate overlap (Fig.~\ref{sky}) highlights an important issue: Given that the SKA,
and two of its pathfinders, will operate in the south, SkyMapper \citep{wol+18} magnitudes should provide the 
optical magnitudes for the 
training of southern sources, upon the data becoming available.\footnote{http://skymapper.anu.edu.au}

\subsection{Contribution of the IR and UV magnitudes}

\subsubsection{Physical interpretation}
\label{physint}

\begin{figure}
\centering \includegraphics[angle=-90,scale=0.48]{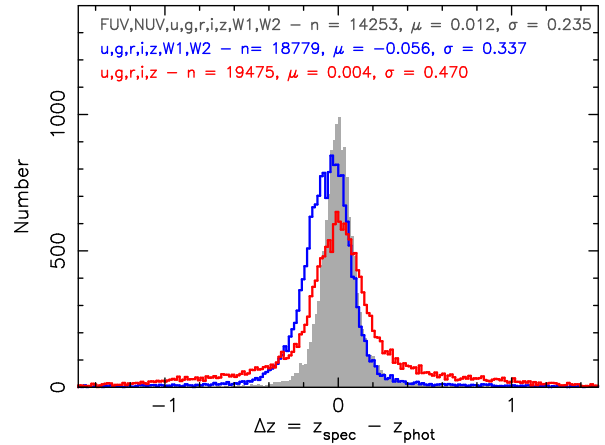}
\caption{The $\Delta z$ distributions using the SDSS colours only, the SDSS+WISE ($W1, W2$)  and the full SDSS
+WISE ($W1, W2$)+GALEX colour set..}
\label{SDSS_histo}
\end{figure}

The improvement in the redshift prediction obtained by the inclusion of the infrared and ultra-violet photometry has
previously been noted by \citet{bmh+12} and \citet{bcd+13}. They attribute the inclusion of these to the breaking of the
redshift degeneracy, which arises when using the $u,g,r,i,z$ photometry alone (e.g. \citealt{rws+01,mhp+12}).  From our
own data, without the full photometry we see that the results degrade significantly
(Fig.~\ref{SDSS_histo}), confirming the importance of the IR and UV data.

\begin{figure}
\centering \includegraphics[angle=-90,scale=0.48]{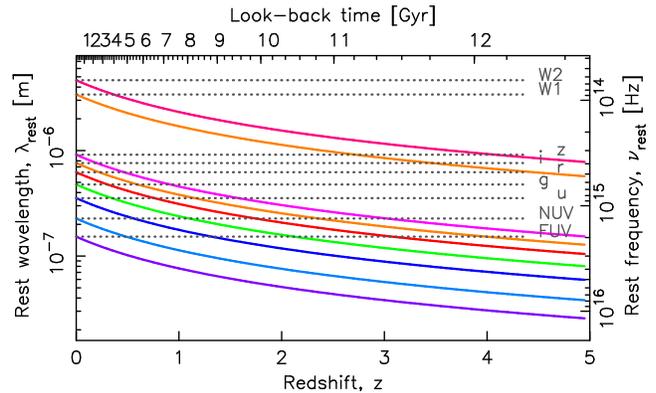}
\caption{The dependence  of source-frame wavelength with redshift for our photometric bands. The dotted lines
on the left-hand ordinate show the observed-frame values and the curves branching off from these the rest-frame wavelength 
as a function of source redshift.}
\label{evolv}
\end{figure}

\citet{cm19} found an  empirical  relationship between the redshift and the ratio of the $U-K$ and $W2-FUV$ colours
in the {\em rest-frame} of the source. 
They hypothesised that this was due to the UV--NIR colours tracing the luminosity of the active galactic nucleus and, as seen
from Fig.~\ref{evolv},
the observed colours will depend strongly upon the redshift of the source. For instance, to measure the $FUV$ and $i$
luminosities at $z=3$ requires the $r$ and $W1$ magnitudes in the observed-frame. 
As noted by \citet{cur20}, inclusion of the mid-infrared $W3$ \& $W4$ magnitudes to
the SDSS training  has little effect on $\sigma_{\Delta z}$, which we confirm is also the case for the DL algorithm.\footnote{Their inclusion reduces the validation sample size from
14\,253 to 11\,045, but with no reduction in $\sigma_{\Delta z}$.} 

\citet{cur20} also noted the  $\lambda\sim1~\mu$m inflection in the SED (e.g. \citealt{em86,bar87,ewm+94}),
which was attributed to 
$\lambda\gapp1~\mu$m  NIR emission from heated
dust. An alternative  explanation is that this is due to H$\alpha$ ($\lambda = 656$~nm) emission, which, at $\lambda =
656$~nm, would be apparent between the $r$ and $i$ bands at $z=0$, shifting to $W1$ at $z\gapp4$.  However, as per
\citet{cur20}, we see little evidence of the shifting of this feature in either of the SEDs (Figs.~\ref{SED}, \ref{OSED}
\& \ref{LSED}), although there does appear to be inflections at $\approx3~\mu$m and 
$\approx0.6~\mu$m in the SDSS and related (LARGESS \& FIRST) spectra (Fig.~\ref{mean_SEDs}). 
A $\approx3~\mu$m feature has been previously noted in type-2 objects \citep{hmg+17} and so we attribute the
$\approx0.6~\mu$m inflection to the rise of the ``big blue bump'', due to thermal emission from the accretion disk.

\begin{figure}
\centering \includegraphics[angle=-90,scale=0.52]{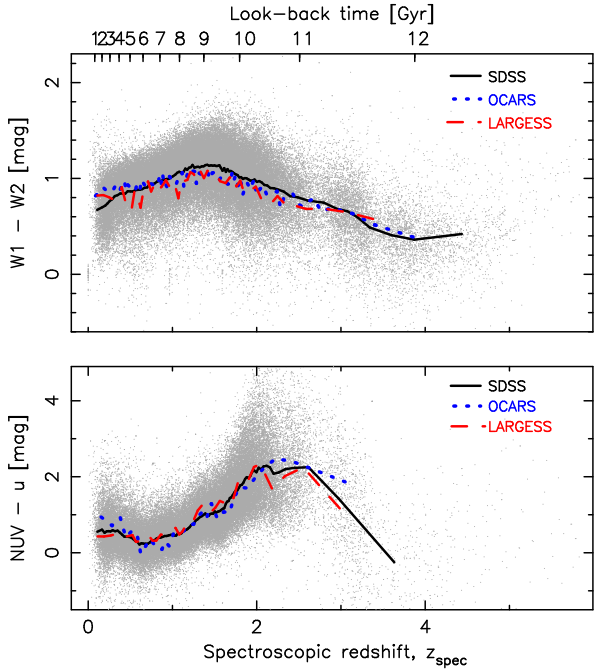}
\caption{Top: The $W1-W2$ colour versus redshift (cf. \citealt{akb+10,rmb+15}). Bottom: The $NUV-u$ colour versus
  redshift.  In both panels the points show the SDSS QSOs of the training sample and the traces show the average values
  of the three main samples.}
 \label{2-colour}
\end{figure}

Any inflection in Figs.~\ref{SED}, \ref{OSED}
\& \ref{LSED} could,  of course,  be somewhat masked by the limited $J,H,K$ photometry of our sample (see
Sect.~\ref{cur_lim}), although the $W1-W2$ colour does exhibit a clear redshift dependence for all samples 
(Fig.~\ref{2-colour}, top), 
where the peak at $z\sim1.5$ corresponds to the rest-frame $J,H,K$ bands.  We also see that, despite our reservation of
the applicability of the SDSS training to radio-loud sources (Sect.~\ref{rfd}), both OCARS and LARGESS follow a similar
evolution to the SDSS QSOs.  This also applies at the other wavelength extreme, where the evolution of $NUV-u$ colour is
also similar between the three main samples (Fig.~\ref{2-colour}, bottom)\footnote{Given our suspicions about the $FUV$
  photometry, especially at high redshift (Figs.~\ref{SED}, \ref{OSED} \& \ref{LSED}), we choose this colour rather than
  $FUV-NUV$.}, all exhibiting a steep evolution in $NUV-u$ at $z\gapp1$ due to the Lyman-break.\footnote{For the evolution
  of the constituent magnitudes see Fig.~\ref{mags-z}.}

In the case of continuum emission from heated dust, we would expect an increase in redshift (and thus, luminosity) to
cause an increase in the peak frequency of the modified black body emission \citep{cd18}. This increase would counteract
any redshift in the peak of the profile, curtailing any apparent shift in the NIR peak perhaps holding the observed NIR
peaks at close to $\lambda\sim1~\mu$m in these relatively low resolution SEDs.
 
Lastly, the $W1-W2$ colour is hypothesised to trace hot dust, from
AGN activity, and the $W2-W3$ colour warm dust, from star formation \citep{jcm+11,dyt+12}.
The fact that the $W1-W2$ colour
exhibits a tighter correlation with source redshift \citep{cd18} suggests that, due to the Malmquist bias,
higher AGN activity is being detected at higher redshift, thus contributing to the determination
of the photometric redshifts \citep{cm19}. Therefore, the MIR 
magnitudes should  also be useful in tracing rest-frame  NIR ($W1$ \& $W2$). However, this would
not take effect until $z\gapp5$. 

\subsubsection{Current limitations} 
\label{cur_lim}

In order to investigate the source of the inaccuracies in $z_{\rm phot}$, we perform a 10-dimensional linear regression of the magnitudes and spectroscopic
redshift versus $\Delta z$ (Table~\ref{T3}).
\begin{table}  
\centering
 \caption{Results of the 10-dimensional linear regression of the standardised ($\mu=0, \sigma=1$) magnitudes
and $z_{\rm spec}$ versus $\Delta z$ for the SDSS test data. $m$ is the gradient, followed by the standard error. The intercept has a value of $c=0.0371\pm0.002$. The final column gives
 the $t$-statistic. }
\begin{tabular}{@{}l  r l r @{}}
\hline
\smallskip
Parameter & $m$ & $\sigma_m/\sqrt{n}$ & $t$ [$\sigma$] \\
\hline
$FUV$ &   0.0222  &   0.002   &     9.115  \\  
$NUV$ &  -0.1490   &  0.003  &    -44.995   \\   
$u$ &     -0.0728 &    0.005 &     -15.964  \\
$g$ &   0.1054 &  0.009  &      12.398     \\
$r$ &      0.1693  &   0.011  &     16.093   \\   
$i$ &    0.1013  &  0.010    &    9.662    \\  
$z$ &     -0.1595 &    0.008  &   -19.877 \\     
$W1$ &   -0.1643   &   0.007   &   -25.177 \\  
$W2$ &       0.0040  &    0.007  &      0.608   \\  
$z_{\rm spec}$ &   0.2236&     0.003  &   80.920  \\
\hline
\end{tabular}
\label{T3}  
\end{table} 
Although, most of the parameters are significant ($|t|\gapp3\sigma$), as discussed above, the NIR ($W1$) and
UV ($NUV$) appear to be extremely important. 

We also note the high significance of $z_{\rm spec}$ and,
\begin{figure}
\centering \includegraphics[angle=-90,scale=0.48]{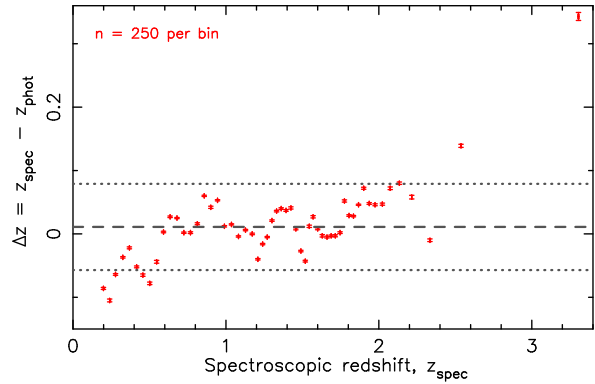}
\caption{$\Delta z$ versus $z_{\rm spec}$ for the SDSS test data binned into $50$ bins of 250 QSOs
each (leaving 3 remaining). The error bars show the
standard error, $\sigma/\sqrt{n}$, and the dashed and dotted lines show the mean and standard deviation of the 
binned values, respectively.} 
\label{dz}
\end{figure} 
plotting this  against $\Delta z$ (Fig.~\ref{dz}), we see that the positive $\Delta z$ values at $z_{\rm spec}\gapp2$
are more extreme than the negative ones at $z_{\rm spec}\lapp1$. Referring to Fig.~\ref{evolv}, at $z_{\rm spec}\sim2$
only the bluer rest-frame bands ($FUV$ to $u$) are observed (over $g$ to $z$), with no coverage until
$z$ in the rest-frame ($W1$ observed). That is, the large gap between the observed $z$ and $W1$ bands, 
manifests in a loss of crucial data ($g,r,i$) in the rest-frame, leading to the drop in accuracy at $z_{\rm spec}\gapp2$.

This gap could potentially be covered with the inclusion of the 2MASS $J,H,K$ bands (see Fig.~\ref{SED}), which were
included in our mining of the photometry (Sect.~\ref{sec:data}), cf.  the UKIDSS $Y,J,H,K$ bands included by
\citet{bcd+13}, Sect. ~\ref{cwpm}. However, the inclusion of these bands reduces the sample to just
19\,077
QSOs. Using an 80:20 training:testing split, we obtain very similar results to our standard $u,g,r,i,z,W1,W2,NUV,FUV$
model, but with the validation sample reduced from 14\,253 to 3815.  
Note that the implementation of a similarly wide array of bands has been used to yield photometric redshifts
through template fitting, specifically by \citet{asu+17} who obtain $\sigma_{\text{NMAD}} = 0.06$ for 5961 X-ray sources
in Stripe 82.

\subsubsection{Excluding the blue and UV data}
\label{exbu}

\begin{figure}
\centering \includegraphics[angle=-90,scale=0.52]{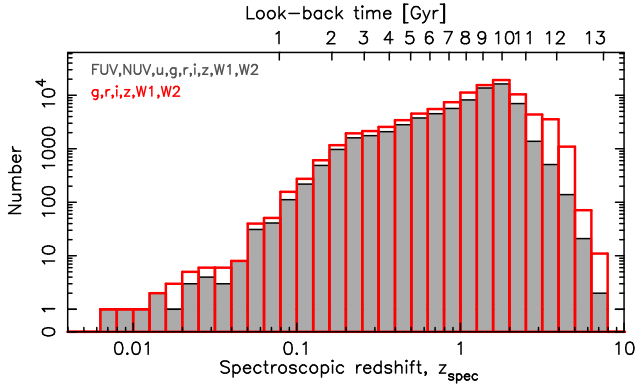}
\caption{The redshift distribution for the SDSS QSOs with all nine magnitudes 
(filled -- 71\,267 sources) and those with $g,r,i,z,W1,W2$ magnitudes (unfilled -- 95\,351 sources). 
It is seen that the additional 23\,084 sources mostly have redshifts $z\gapp2$.}
\label{hist_noblue}
\end{figure}

Given the  disadvantage of requiring all nine magnitudes and the fact that the ``bluer'' bands ($u,NUV,FUV$) are
more likely to be undetected, particularly at high redshift (Fig.~\ref{hist_noblue}), we investigate the effect
of removing these magnitudes from the DL model.
Excluding these gives a matching coincidence of  95\%, cf. 71\%, of the original sample (Sect.~\ref{sec:data}).
As before, we use an 80:20 training--validation split on the SDSS data, giving 
a validation sample of 19\,070 QSOs with the $g,r,i,z,W1,W2$ magnitudes.
\begin{figure}
\centering \includegraphics[angle=-90,scale=0.42]{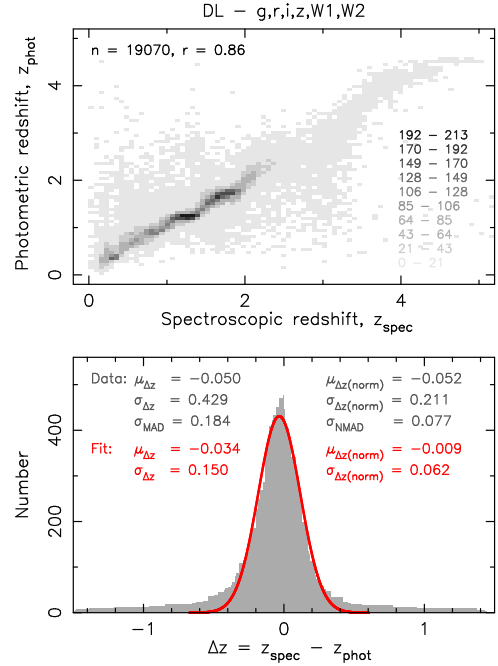}
\caption{As Fig.~\ref{f1} (right) -- the predictions based on the DL method on the SDSS sample, but excluding
  the colours which use  $u$ and GALEX magnitudes.}
\label{no_blue}
\end{figure} 
From Fig.~\ref{no_blue}, we see a considerable spread in $\Delta z$ ($\sigma_{\Delta z}[\text{data}]=0.429$)
compared to the full $FUV,NUV,u,g,r,i,z,W1,W2$ complement ($\sigma_{\Delta z}[\text{data}]=0.235$).
\begin{figure}
\centering \includegraphics[angle=-90,scale=0.48]{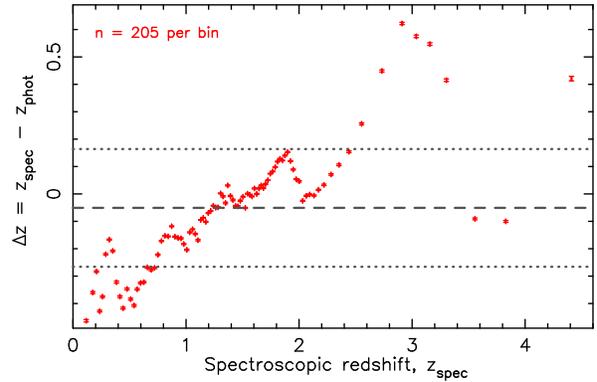}
\caption{As Fig.~\ref{dz}, but excluding the "bluer" ($u, NUV, FUV$) magnitudes and binned into $93$ bins of 205 QSOs
each (leaving 5 remaining). 
The error bars show the
standard error $\sigma/\sqrt{n}$ and dashed and dotted lines show the mean and standard deviation of the 
binned values, respectively.} 
\label{dz-noblue}
\end{figure} 
Since the UV magnitudes may be important in tracing the AGN activity at low redshift (Sect.~\ref{physint}), we expect
that most of the uncertainty, due to the omission of the bluer magnitudes, would occur at $z\sim0$.  This is confirmed
in Fig.~\ref{dz-noblue}, where the previous inaccuracy at $z\gapp2$ appears also to be exacerbated
(cf. Fig.~\ref{dz}). In summary, although the exclusion of the colours which use the $u, NUV, FUV$ magnitudes does
increase the number of high redshift detections, this is at the expense of the low redshift accuracy, with the
additional high redshift estimates gained being of relatively low quality.

\begin{figure*}
\centering \includegraphics[angle=-90,scale=0.55]{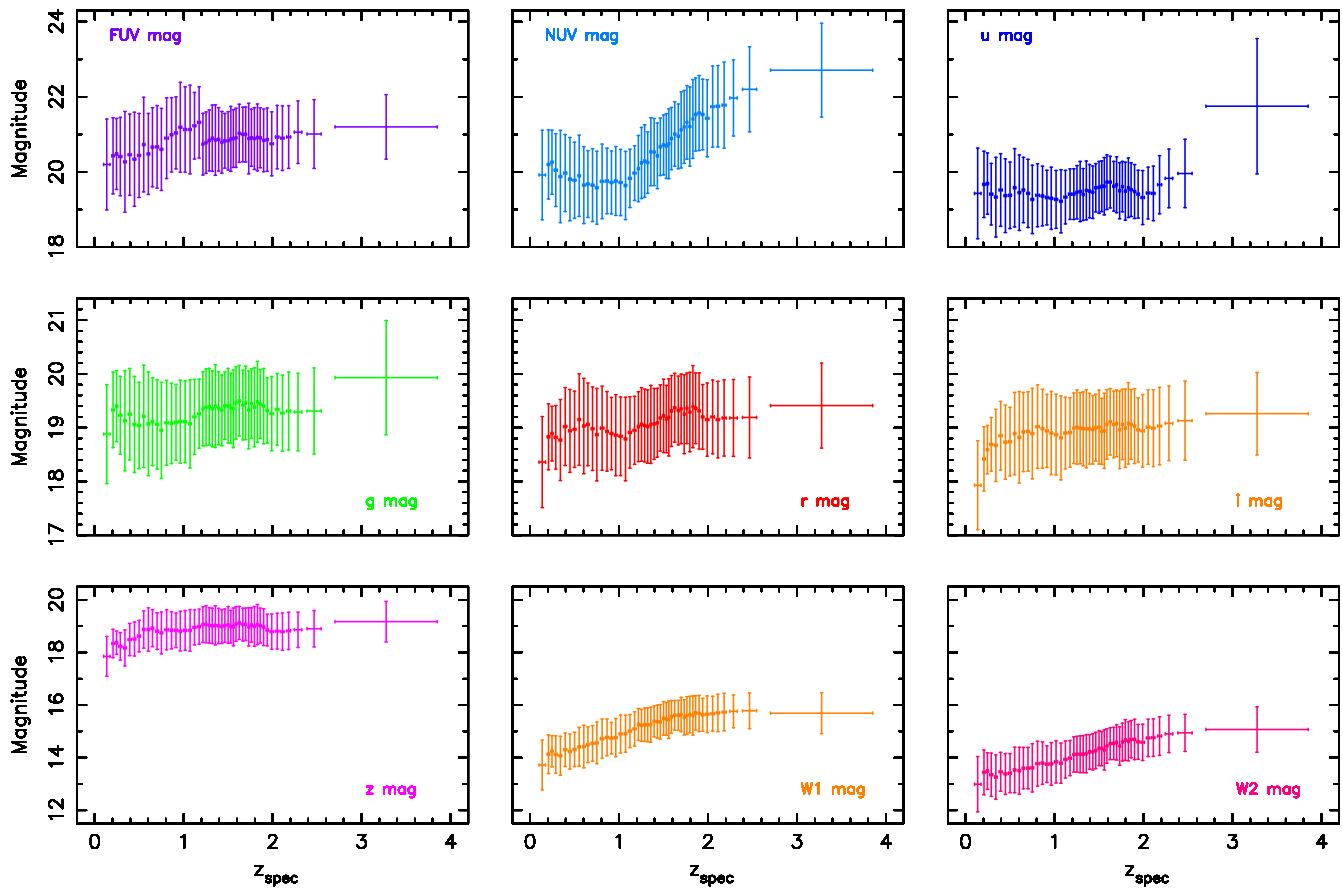} 
\caption{The magnitude versus redshift for the SDSS validation sample in 50 bins of 151 QSOs (leaving one remaining). As
  mentioned above, the high redshift $FUV$ data are probably unreliable.} 
\label{mags-z}
\end{figure*}

The multi-dimensional linear regression of the magnitudes versus $\Delta z$ (Table~\ref{T3}) suggests that both the
$NUV$ and $u$ magnitudes are especially crucial in obtaining a reliable photometric redshift. Examining the redshift
distribution of the magnitudes (Fig.~\ref{mags-z}),
we see that the $NUV$ magnitude climbs rapidly at $z\gapp1$ and the
$u$ at $z\gapp2$, which is consistent with the rest-frame $\lambda= 1216$~\AA\ Lyman-break. The $FUV$
  magnitude also exhibits this at the expected $z\gapp0.3$, although as established from Figs.~\ref{SED}, \ref{OSED}
\& \ref{LSED}, this band is not expected to be reliable at $z\gapp2$. 

The poorer performance resulting from the exclusion of the bluer magnitudes, suggests that the algorithm relies heavily
upon the Lyman-break in estimating the redshift.  The fact that both of these magnitudes peter out (at $NUV\gapp22$ \&
$u\gapp20$) is most likely responsible for the poor photometric redshifts estimates at high redshift.  Improving upon
this would require an automated and reliable method of including the non-detections or targetted longer integrations
upon the sources of interest.  Nevertheless, from Fig.~\ref{dz}, it appears that the photometric redshifts are
statistically accurate up to $z\sim2.5$, which, at a look-back time of 11~Gyr, covers the past 80\% of cosmic history.

\section{Conclusions}

Forthcoming radio continuum surveys on the next generation of telescopes are expected to yield vast numbers of sources for
which the redshifts will be unknown. Since these are observationally expensive to obtain, as well as introducing a bias
towards the most optically bright objects, a rapid method of estimating accurate redshifts from the 
photometry would vastly increase the scientific value of these surveys.

Given the large datasets involved, machine learning is the most promising means.  While there has been success training
the algorithms on optical (SDSS) photometry alone, we have previously shown that the combination of SDSS, WISE $W1$ \&
$W2$ and GALEX colours leads to a significant increase in accuracy \citep{cur20}. Here we compare our previous method,
the $k$-Nearest Neighbour, with the Decision Tree Regression and Deep Learning methods.  Mining the
$FUV,NUV,u,g,r,i,z,W1,W2$ photometry of the 100\,000 QSOs selected from the SDSS DR12, gives 71\,267 sources which are
detected in all nine bands. Testing the full sample and various sub-samples shows the DL algorithm to perform the best,
as measured by the (normalised) standard deviation, the (normalised) median absolute deviation, the regression
coefficient and the gradient of the linear fit between the predicted and measured redshifts. Training the DL algorithm on
80\% of the SDSS sample and validating on the other 20\%, yields an accuracy of $\Delta z < 0.1$ up to $z\sim2.5$ 
which corresponds to look-back times of 11 Gyr.

In order to determine the suitability of the DL model in predicting redshifts for other, radio-selected, samples, as for
the SDSS sample, we scrape the photometry from various databases and bin these into the appropriate bands.  We
then use our full SDSS DR12 sample to train the model and validate this on the other catalogues. We find this
cross-training to be successful, yielding photometric redshifts up to $z\sim4$, with a standard deviation in $\Delta z$
of $\sigma_{\Delta z}[\text{data}] \approx 0.25 - 0.37$.  This is despite the mean radio flux densities of the samples
differing by two orders of magnitude from the training sample, as well as there being a clear difference in the mean
optical SEDs (cf. \citealt{ewm+94}).

As per the kNN and other methods \citep{bmh+12,bcd+13}, the accuracy of the photometric redshift predictions depends
heavily upon the addition of the infrared and ultra-violet photometry to the standard $u,g,r,i,z$
(\citealt{rws+01,wrs+04,bbm+08,mhp+12,hdzz16}). As noted by \citet{cm19}, over a large range of redshifts, a given
rest-frame magnitude could occur in several other bands.  This suggests that the mid-infrared $W3$ \& $W4$ bands may be
important at $z\gapp5$, where the data are currently sparse, and that the current loss in accuracy at $z\gapp2$ is due
to the large gap between the SDSS and WISE bands.  This could be somewhat remedied with the inclusion of the NIR $J,H,K$
bands \citep{bcd+13}, although the large reduction in sample size prevents this being the case for our sample.

Our data scraping and binning of the photometry has the advantage over other methods which use the SDSS magnitudes
directly in that it produces a model that can be used to train other samples which may have little SDSS, but other
optical photometry available (e.g. $B,G,V,R$).  Our DL method has the advantage over similar applications of neural
networks in that it utilises an off-the-shelf deep learning library with basic hyperparameters -- two {\em ReLu} layers
and one {\em tanh} layer each comprising 200 neurons. It also runs rapidly on a standard laptop computer, compared to
the DCMDN method \citep{dp18}, which requires a cluster, and the MLPQNA method \citep{bcd+13}, which requires the prior
``pruning'' of features in order to speed up the computation.  The main disadvantage, common to these other methods, is
the requirement of the detection of the source over nine different observing bands, particularly the GALEX bands which
are the most restrictive. Furthermore, the requirment of SDSS magnitudes to validate the all-sky surveys halves the
number of sources which can be used.  This highlights the need for southern sky training data (e.g. using SkyMapper), if
the aim is to obtain photometric redshifts for continuum sources detected with the SKA.

\section*{Acknowledgements}

We wish to thank the referee for their very helpful comments.
This research has made use of the NASA/IPAC Extragalactic
Database (NED) which is operated by the Jet Propulsion Laboratory, California Institute of Technology, under contract
with the National Aeronautics and Space Administration and NASA's Astrophysics Data System Bibliographic Service. This
research has also made use of NASA's Astrophysics Data System Bibliographic Service.
Funding for the SDSS has been provided by the Alfred P. Sloan Foundation, the Participating Institutions, the National Science Foundation, the U.S. Department of Energy, the National Aeronautics and Space Administration, the Japanese Monbukagakusho, the Max Planck Society, and the Higher Education Funding Council for England. 
 This publication makes use of data products from the Wide-field Infrared Survey
Explorer, which is a joint project of the University of California, Los Angeles, and the Jet Propulsion
Laboratory/California Institute of Technology, funded by the National Aeronautics and Space Administration.  This
publication makes use of data products from the Two Micron All Sky Survey, which is a joint project of the University of
Massachusetts and the Infrared Processing and Analysis Center/California Institute of Technology, funded by the National
Aeronautics and Space Administration and the National Science Foundation. GALEX is operated for NASA by the California
Institute of Technology under NASA contract NAS5-98034.  

\section*{Data availability} 

Data and SDSS {\sf TensorFlow} training model available on request.


\label{lastpage}

\end{document}